\documentclass[5p,twocolumn]{elsarticle}
\usepackage{amsmath,amsfonts,amssymb,bm}
\usepackage{graphicx}
\usepackage{lipsum}

\newcommand{\abs}[1]{\ensuremath{|#1|}}
\newcommand{\tr}{\ensuremath{\operatorname{tr}}}
\newcommand{\ket}[1]{\ensuremath{|#1\rangle}}
\newcommand{\bra}[1]{\ensuremath{\langle #1|}}
\newcommand{\ketbra}[2]{\ensuremath{\ket{#1}\!\bra{#2}}}
\newcommand{\sgn}{\ensuremath{\operatorname{sgn}}}
\newcommand{\up}{{\uparrow}}
\newcommand{\down}{{\downarrow}}

\journal{}
\begin{document}
	\begin{frontmatter}
		\title{Heat pump driven by the shot noise of a tunnel contact}
		\author[ICMMaddress]{Robert Hussein}
		\author[ICMMaddress]{Sigmund Kohler}
		\author[UCMaddress]{Fernando Sols\corref{mycorrespondingauthor}}
		\cortext[mycorrespondingauthor]{Corresponding author}
		\ead{f.sols@fis.ucm.es}
		\address[ICMMaddress]{Instituto de Ciencia de Materiales de Madrid, CSIC, Cantoblanco, E-28049 Madrid, Spain}
		\address[UCMaddress]{Departamento de F\'isica de Materiales, Facultad de Ciencias F\'isicas, Universidad Complutense de Madrid, E-28040 Madrid, Spain}
		\begin{abstract}
			We investigate a mechanism for cooling a lead based on a process that replaces
			hot electrons by cold ones.  The central idea is that a double quantum dot
			with an inhomogeneous Zeeman splitting acts as energy filter for the
			transported electrons.  The setup is such that hot electrons with spin
			up are removed, while cold electrons with spin down are added.  The required
			non-equilibrium condition is provided by the capacitive coupling of one
			quantum dot to the shot noise of a strongly biased quantum point contact in the tunnelling limit.
			Special attention is paid to the identification of an operating regime in which the
			net electrical current vanishes.
		\end{abstract}
	\end{frontmatter}
	\section{Introduction}
	
	The program of devising novel refrigeration schemes for electronic systems defines
	a variety of fundamental problems of potential technological interest.
	The pumping of heat from a cold to a hot electron reservoir by suitably manipulating the electron transport may be viewed as a form of rectified motion \cite{BuettikerZPB1987a,Reimann2002a,Hanggi2009a}. Electron pumps can be adiabatic \cite{HumphreyPRL2002a,AvronJSP2004a,PekolaPRL2007a} or (if they use inelastic transitions) nonadiabatic \cite{Rey2007a,ChiJP2012a}. The latter case represents a particular instance of ac-driven electron heat transport \cite{Moskalets2002a,Arrachea2007a}. It has been recognized that the effect of nonadiabatic driving can be very similar to that of an energetic, far-from-equilibrium energy source that promotes inelastic transmission of electrons through a spatially asymmetric setup \cite{PekolaPRL2007b,SegalPRL2008a,SolsPhysicaE2010a,CleurenPRL2012a}.
	In this paper, we follow the approach of these works and investigate the possible use of the shot noise of a nearby quantum point contact (QPC) as the nonequilibrium energy source that, through capacitive coupling, induces rectification in a spatially asymmetric device which in our case will be a double quantum dot (DQD) system.
	
	In mesoscopic physics, recent experimental progress in
	energy harvesting \cite{Radousky2012a} turned the fundamental issue of heat
	balance into a topic of practical interest \cite{Sothmann2015a}.
	In particular, quantum dot setups have been argued to provide a convenient framework for the controlled transport of heat between electron reservoirs \cite{Rey2007a,SegalPRL2008a,SanchezNJP2013a,VenturelliPRL2013a}, as has been experimentally confirmed
	\cite{ThierschmannNJP2013a}. On the other hand, it has also been argued that capacitive coupling to systems with fluctuating charges can be a source of energy and of induced heat transport \cite{SanchezPRB2011a}, a claim that has been recently observed \cite{ThierschmannNatureNInPress2015a}. A similar prediction for quantum cavity systems \cite{SothmannPRB2012a} has also received experimental confirmation \cite{RocheNatureC2015a,HartmannPRL2015a}. Analogous physics has been predicted for small electron systems interacting through the exchange of phonons \cite{Entin-WohlmanPRB2010a} or microwave photons \cite{RuokolaPRB2012a}
	
	When considering electron exchange between reservoirs, one typically
	assumes each lead in a grand canonical state determined by its chemical
	potential and its temperature.  Then thermal excitations involve
	occupied states above the Fermi energy and unoccupied states below.  Our
	focus lies on the associated excitation energies.  Consequently, we speak of
	cooling when electrons in states above the Fermi energy are removed or
	when holes below are filled.  Thus, one way to cool a lead is to
	contact it with another lead at a different chemical potential while applying an
	appropriate energy filter which can be realized, e.g., by the gap of a superconductor
	\cite{Pekola2005a} or by a resonant level. The same principle can be applied to the cooling of
	of a two-dimensional electron gas mediating the charge flow between two leads, as has been
	predicted \cite{EdwardsPRB1995a} and observed \cite{PrancePRL2009a}; for a review
	see \cite{GiazottoRMP2006a}. A recent experiment is presented in Ref.~\cite{KoskiArXive2015a}, which is based on the mechanism proposed in Ref.~\cite{SanchezPRB2011a}.

	The above cooling scheme, however, relies on an applied bias which also causes a net
	electron transfer.  In particular when one lead is actually a small grain,
	one soon reaches a situation in which the grain becomes electrically
	charged and the cooling process comes to an end. In this article we follow Ref.
	\cite{Rey2007a} by proposing
	a mesoscopic heat pump that avoids a net charge
	transport while operating between two leads with equal chemical potentials.
	In consistency with the second law of thermodynamics, such a heat flow
	from cold to hot requires some non-equilibrium condition.  In
	Ref.~\cite{Rey2007a}, this has been theoretically achieved by an ac gating.  Here by
	contrast, we employ the shot noise of a nearby quantum point contact.
	The shot noise of a nearby conductor has already been investigated as a source of rectification
	~\cite{LevchenkoPRL2008a,SanchezPRL2010a,Hussein2015a}.
	Here we adapt the study in Ref. \cite{Hussein2015a}, where the fluctuating conductors is a QPC in the tunnel limit, to the
	pumping of heat with zero net electric current between the cold and the hot leads.
	The advantage of using a charge-fluctuating QPC is two-fold: First, its broad-band excitation is less
	sensitive to small detunings.  Second, most recent quantum dots already
	include such a QPC \cite{Ihn2009a, Taubert2011a} so that our proposal can
	be realized readily.
	
	The behavior of quantum dot systems subject to the effect of nearby fluctuations can be studied as a particular case of the ``particle-bath'' problem, where attention is paid to the effect on a few physical variables (or on a reduced Hilbert space) of many degrees of freedom that are mathematically traced out \cite{Caldeira1983a}. The effect of the dissipative environment is that of rendering the dynamics effectively irreversible despite the time-reversal symmetry of the underlying microscopic dynamics.  On a macroscopic level, irreversibility
	becomes manifest in the fact that heat can flow spontaneously only from a hot to
	a cold reservoir. Such a quantum dissipation approach underlies the master equation calculation described in section 2, where the leads degrees of freedom are traced out.

	This paper is arranged as follows. In section 2 we describe the model considered and the calculation method employed. Section 3 contains analytical and numerical results for the case of noninteracting electrons. In section 4, we discuss the effect of Coulomb repulsion at the dots, focusing on the displacement of the particle-hole symmetry point. Section 5 is devoted to conclusions. Appendix A provides a detailed account of the arguments leading to the analytical results obtained for the charge and heat currents which are presented in section 3.
	
	\begin{figure}
		\centerline{%
		\includegraphics[]{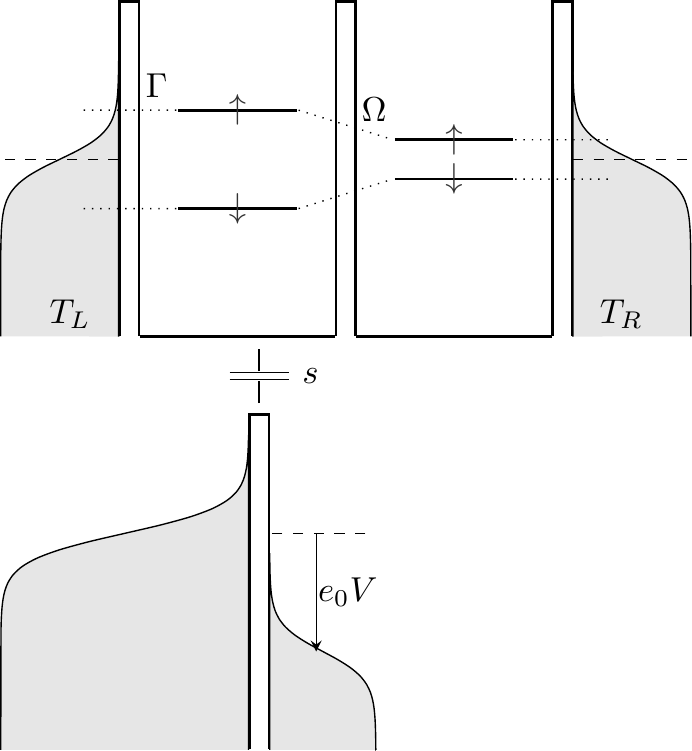}}
		\caption{\label{fig:model}
		Sketch of the double quantum dot (DQD) coupled to a QPC in the tunnel regime.  We focus on the
		heat balance in the right lead of the DQD, where hot spin-up electrons are removed
		while cold spin-down electrons are added.  The coupling to the shot noise
		of the QPC creates a non-equilibrium situation that breaks the symmetry
		between the forward and the backward process.  In the absence of the QPC,
		detailed balance inhibits cooling.}
	\end{figure}
	
	\section{Model and master equation}
	
	\subsection{DQD coupled to a QPC in the tunnel regime}
	
	Our model sketched in Fig.~\ref{fig:model} consists of two capacitively
	coupled electric circuits, namely an unbiased DQD and a strongly biased
	tunnel contact.  The latter entails non-equilibrium noise on the former
	and, thus, drives the DQD out of equilibrium.  The DQD is modeled by the
	Hamiltonian
	\begin{equation}
		\begin{split}
			\label{Hdqd}
			H_\text{DQD} ={}&
			\frac{\Omega}{2} \sum_{\sigma=\up,\down}( c_{1\sigma}^\dagger
			c_{2\sigma}+c_{2\sigma}^\dagger c_{1\sigma})
			+ \frac{U}{2} \sum_{\kappa\neq\kappa'} n_\kappa n_{\kappa'}
			\\{}&
			+ \sum_{i=1,2} B_i(c^\dagger_{i\up}c_{i\up}-c^\dagger_{i\down}c_{i\down})
			+ V_\text{gate}\sum_{\kappa}c_\kappa^\dagger c_\kappa
		\end{split}
	\end{equation}
	where $i=1,2$ and $\sigma=\up,\down$ label the two quantum dots and the
	spin degree of freedom, respectively, while the multi index $\kappa = (i,
	\sigma)$ combines both.  The terms describe spin-independent tunneling
	$\Omega$, Coulomb repulsion $U$, an inhomogeneous Zeeman splitting $B_i>0$,
	and a global shift of the onsite energies caused by a background gate
	voltage (with the sign convention that a positive $V_\text{gate}$ shifts the DQD
	levels upwards).  For simplicity, we restrict ourselves to undetuned dot levels (in
	the absence of the magnetic field) and assume that both intra-dot and
	inter-dot Coulomb repulsion are equal.
	
	Dot~1 is tunnel coupled to the left lead $L$ described by $H_\text{dot-lead} =
	\sum_{q,\sigma} \epsilon_q (c^\dagger_{L,q,\sigma}c_{1,\sigma} +c_{1,\sigma}^\dagger
	c_{L,q,\sigma})$ and dot~2 accordingly to the right lead.  The effective
	dot-lead coupling is given by the rate $\Gamma = 2\pi\sum_{q} |V_{L,q}|^2
	\delta(\epsilon-\epsilon_q)$ which we assume independent of the energy
	$\epsilon$ and the spin projection $\sigma$.  Then the equilibrium distribution of the
	lead electrons is given by the Fermi function $f_\ell(\epsilon) =
	[\exp(\epsilon/k_BT_\ell)+1]^{-1}$ where $\ell=L,R$, i.e., while setting
	$\mu=0$ for both leads, we allow for different lead temperatures
	
	Our second system is a QPC in the tunnel limit between two leads modeled by the Hamiltonian
	$H_\text{QPC} = \sum_k \epsilon_k c_k^\dagger c_k + \sum_{k'} \epsilon_{k'}
	c_{k'}^\dagger c_{k'}$, where $k$ and $k'$ label the modes of the left and
	the right lead, respectively, including the spin.  The leads are weakly
	coupled by the tunnel Hamiltonian $\Lambda = \Lambda_++\Lambda_-$, where
	\begin{equation}
		\label{HQPC}
		\Lambda_+ = \sum_{k,k'} t_{kk'} c_{k'}^\dagger c_k ,
	\end{equation}
	transfers an electron from the left to the right QPC lead, while
	$\Lambda_-\equiv \Lambda_+^\dagger$ describes the opposite process.
	In the continuum limit, the matrix elements $t_{kk'}$ are encompassed by
	the energy-independent QPC conductance $G = 2\pi \sum_{kk'}
	|t_{kk'}|^2 \delta(\epsilon-\epsilon_k) \delta(\epsilon-\epsilon_{k'})$
	in units of the conductance quantum $G_0 = e_0^2/h$.
	The electrons on dot~1 enhance, via Coulomb repulsion, the barrier
	between the QPC leads and thereby reduce the tunnel matrix elements $t_{kk'}$.
	This effect is captured by a prefactor $x = (1-s n_{1})$
	in the tunnel Hamiltonian such that
	\begin{equation}
		\label{Hqpc}
		H_\text{QPC}^\text{tun}
		= x(\Lambda_++\Lambda_-)
	\end{equation}
	accounts for both the QPC and its coupling to the DQD.  The latter is
	mediated by the occupation of dot~1, $n_1 = \sum_\sigma n_{1\sigma}$.  For
	consistency, the dimensionless coupling $s$ must obey $0\leq s\leq1/2$.
	
	\subsection{Master equation}
	
	Our theoretical description is based on the formal elimination of all four
	leads such that we remain with a reduced master equation for the DQD.
	While the treatment of the leads coupled to the DQD follows a standard
	procedure, the systematic elimination of the tunnel contact is less common
	and has been performed only recently \cite{HusseinPRB2014b, Hussein2015a}.
	In those works, a counting variable for the tunnel contact allowed to compute
	the full counting statistics and correlation functions, while here it is
	sufficient to compute the action of the QPC on the DQD.  We sketch here the
	derivation of the formalism of Ref.~\cite{HusseinPRB2014b, Hussein2015a} as
	required for our present purposes.
	
	We start from the Liouville-von Neumann equation for the full density
	operator which we transform to the interaction
	picture with respect to $H_\text{DQD}$ and the lead Hamiltonians.  The
	remaining terms are treated within second-order perturbation theory in the
	dot-lead tunnelings and in the QPC tunneling to
	obtain the Bloch-Redfield master equation \cite{Redfield1957a}
	\begin{align}
\dot \rho = -\frac{i}{\hbar}[H_S,\rho]
-\frac{1}{\hbar^2}\sum_{n}\int\limits_{0}^{\infty} dt\,
\tr_{\textrm{leads}} [V_n, [\tilde V_{n}(-t), \rho\otimes R_0]],
\label{BRME}
	\end{align}
	for the reduced DQD density operator $\rho$.  $R_0$ refers to the grand
	canonical ensemble of each lead, while the operators $V_n$ represent
	$H_\text{QPC}^\text{tun}$ and the two tunnel contributions in
	$H_\text{DQD-leads}$. Here, $\tilde V_{n}(t)$ stands for $V_{n}$ in the interaction picture.
	
	The evaluation of the Liouvillian $\mathcal{L}_\text{DQD-leads}$ for the
	incoherent DQD-lead tunneling is rather standard, see e.g.\ the appendix of
	Ref.~\cite{Hussein2012a}.  It yields operators that describe jumps
	between many-particle DQD states differing by one electron.  The transition
	rates contain Fermi functions reflecting the initial occupation of
	the lead modes.  To obtain an expression for the heat balance, we multiply in the
	superoperator for the electric current each tunnel process by the energy with
	respect to the chemical potential which the electron carries to the lead \cite{SivanPRB1986a,Moskalets2002a,Rey2007a,Arrachea2007a}.
	
	To obtain the Liouvillian for the action of the QPC, we evaluate the
	$t$-integral in Eq.~\eqref{BRME} and find the QPC Liouvillian
	\begin{equation}
		\label{Lqpc0}
		\begin{split}
			\mathcal{L}_\text{QPC}\rho =
			\frac{1}{2\hbar^2}\int_{-\infty}^{+\infty} dt\; C(t)\big[
			& \tilde x(-t)\rho x + x\rho\tilde x(t)\\
			-& x\tilde x(-t)\rho-\rho\tilde x(t) x
			\big] .
		\end{split}
	\end{equation}
	Symmetrizing the time integral amounts to neglecting the energy
	renormalization stemming from principal values.  A main ingredient is the
	correlation function of the QPC tunnel operator, $C(t) = \langle
	\Lambda(t)\Lambda(0)\rangle = C_+(t)+C_-(t)$, where $C_\pm(t) =
	\langle\Lambda_\mp(t)\Lambda_\pm(0)\rangle$ is readily evaluated from its
	definition and the assumption that the leads are voltage biased.  In
	Fourier representation it reads \cite{Ingold1992a}
	\begin{equation}
		\label{Comega}
		C_\pm(\omega)
		= G\, \frac{\hbar\omega\pm e_0V}{1-\exp[-(\hbar\omega\pm e_0V)/k_BT]} .
	\end{equation}
	
	We restrict ourselves to the limit in which the QPC bias $V$ is much larger
	than any other relevant frequency scale of the DQD.  Then $C(\omega) = \hbar e_0GV$
	becomes frequency independent so that $C(t) \propto \delta(t)$.  Then we
	obtain for the QPC Liouvillian the Lindblad form
	\begin{equation}
		\mathcal{L}_\text{QPC}\rho =
		\gamma\big( x\rho x  - [x^2,\rho]/2 \big)
	\end{equation}
	with the effective rate $\gamma = 2\pi I_\text{QPC}/e_0$.
	For a treatment beyond the large-bias limit, see Ref.~\cite{Hussein2015a}.
	
	In our numerical implementation of the master equation formalism, we
	use the many-particle eigenstates of the DQD Hamiltonian as a basis and keep
	all off-diagonal elements of the density matrix.  This ensures to capture
	the level repulsion stemming from the tunnel coupling, which is rather
	relevant for levels close to the Fermi energy.
	
	\section{Heat balance in the absence of interaction}
	
	To explain the central idea of the cooling mechanism, we consider the
	channels for the spin-up electrons and for the spin-down electrons
	separately.  To do so, we ignore Coulomb repulsion which couples the
	spin-up and the spin-down channels.  Later we will see that the interaction
	term in the Hamiltonian~\eqref{Hdqd} shifts the working point and affects
	the efficiency.
	
	\subsection{Pumping mechanism}
	
	We consider the DQD sketched in the upper half of Fig.~\ref{fig:model}.
	Let us focus on the channel for the spin-down electrons and assume equal
	temperatures, $T_L=T_R$.  Suppose that an electron enters dot~1 from the
	left lead at the corresponding onsite energy.  The electron may proceed to
	dot~2 and eventually to the right lead.  Due to the lower occupation at
	higher initial energy, the opposite process occurs with lower probability.
	Thus we expect a net transport that in the right lead fills holes below the
	Fermi surface, which corresponds to cooling.
	
	However, this picture is incomplete.  Owing to the tunnel coupling between the
	dots, DQD eigenstates are delocalized, but in such a way that (spin-down) electrons in the ground state
	are mostly in the left dot while the excited ones dwell mostly in the right dot.
	Therefore, electrons in the
	excited states are more likely to leave to the right lead, while electrons
	preferentially enter the ground
	state from the left lead.  The quantitative analysis [see
	Eqs.~\eqref{Ianalyt} and \eqref{Qanalyt} below] reveals that, when both leads 
	have the same temperature and chemical potential, those
	processes fail to yield a net effect, i.e., both the electric current and the
	heat current vanish.  This is of course what one expects for
	such an equilibrium situation.  In more technical terms, the full
	description obeys detailed balance, as it should.
	
	To obtain any net transport, we must drive the system out of equilibrium.  For
	this purpose, we couple the DQD to a strongly biased QPC in the tunnel
	regime; see the Hamiltonian~\eqref{Hqpc}.  Then the shot noise of
	the tunnel current induces transitions between the ground state and the
	excited state of the DQD.  Since the ground state is more strongly populated, we
	witness a net excitation and, consequently, the transport process from left
	to right dominates the one in the opposite direction.  We thus observe both
	charge pumping and cooling of the right lead.
	
	The channel for the spin-up electrons behaves similarly.  It yields a net
	transport of hot electrons from the right to the left lead, which again
	corresponds to cooling.  In the symmetric situation $V_\text{gate}=0$
	sketched in Fig.~\ref{fig:model}, the transported charges of both spin channels
	compensate each other, while the energy transfer adds up.  The net result
	is cooling of the right lead without charge accumulation.
	
	\subsection{Quantitative analysis of the individual channels}
	
	In order to substantiate the above discussion, we have solved the master
	equation for a single spin channel within the rotating-wave approximation.  This
	approximation means that, within a Pauli-type master equation approach,
	off-diagonal density matrix elements are neglected and
	only the populations of the eigenstates are considered.
	The QPC-induced transitions between the (single-particle)
	ground state and the excited state of the DQD have been treated as
	a perturbation.  Moreover, we have ignored the doubly occupied state.
	While details of the calculation can be found in \ref{sec:analytics}, we
	present here the results for the electric current and the heat
	balance in the limit $\gamma\ll\Gamma$:
	\begin{align}
\label{Ianalyt}
I ={}& \frac{\gamma}{4} \frac{f_e-f_g}{1-f_gf_e}
       \frac{\varepsilon\Omega^2}{(\varepsilon^2+\Omega^2)^{3/2}} , \\
\label{Qanalyt}
{\dot Q} ={}& \frac{\gamma}{8} \frac{f_e-f_g}{1-f_gf_e}
    \frac{\Omega^2}{(\varepsilon^2+\Omega^2)^{1/2}} +( V_\text{gate}-\mu) I ,
	\end{align}
	where $\varepsilon$ is the Zeeman energy gradient, $\varepsilon = (B_2-B_1)/2$, and the Fermi
	functions at the QDQ energies are abbreviated as $f_g = f(E_g-\mu)$ and
	$f_e=f(E_e-\mu)$.
	
	If the ground state lies well below the Fermi surface, so that
	$f_g\simeq 1$, while $f_e$ is significantly smaller, the prefactor
	involving the Fermi functions becomes unity.  Then both the electric current
	and the heat balance depend only on the DQD configuration and the distance
	to the symmetry point.
	
	An important implication of Eqs.~\eqref{Ianalyt} and \eqref{Qanalyt} is
	that both vanish in the absence of the QPC, since then $\gamma=0$.  This underlines that our
	treatment is consistent with detailed balance.
	
	\begin{figure}
		\includegraphics[width=0.9\columnwidth]{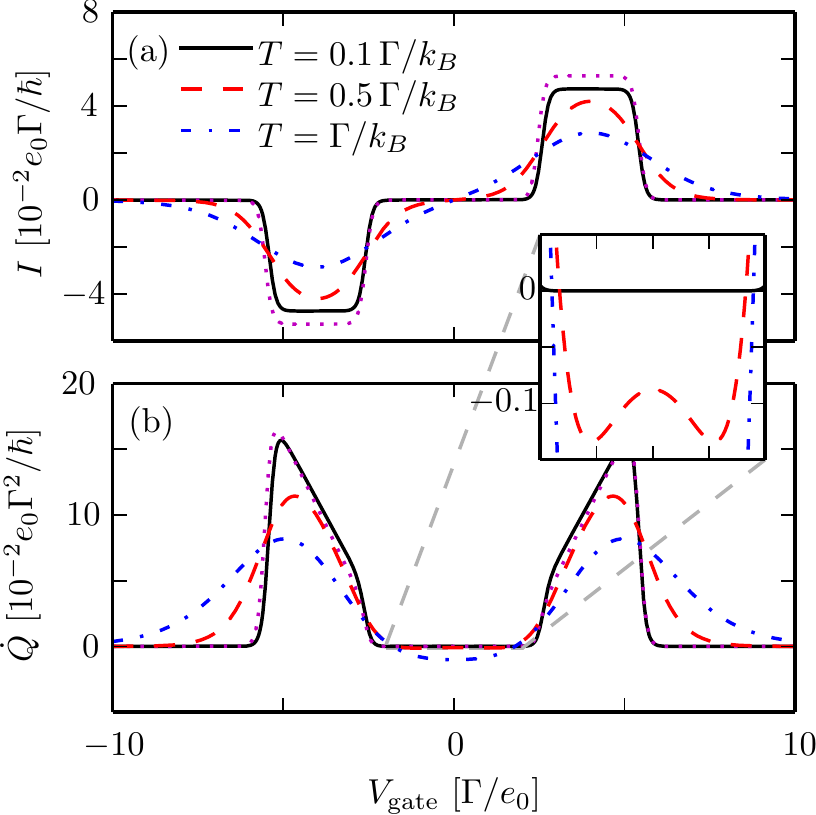}\caption{\label{fig:U0}
		(a) Current and (b) heat balance as a function of the gate voltage
		$V_{\textrm{gate}}$ for various DQD temperatures and zero Coulomb
		interaction. Parameters are $\Gamma=2\Omega=B_1/10=B_2/6=2k_BT_{\textrm{QPC}}=e_0V/80$,
		QPC conductance $G=1$, and coupling strength $s=0.1$.
		The dotted lines mark the analytical solution given in \ref{sec:analytics},
		for $k_BT = \Gamma/10$.
		Inset: Zoom of the heat balance in the region close to $V_\text{gate}=0$
		where ${\dot Q}<0$.}
	\end{figure}
	
	\subsection{Cooling}
	
	Our next goal is to provide numerical results for the different operating
	regimes.  In doing so, we plot in Fig.~\ref{fig:U0} the electric current and the
	heat balance as a function of the gate voltage and for parameters that
	otherwise correspond to the sketch in Fig.~\ref{fig:model}.
	
	For very negative gate voltage, all four levels lie below the Fermi
	surface and no transport occurs.  When shifting all levels upwards, the
	excited state of the spin-up channel will cross the Fermi function and
	a pump current from the right to the left lead sets in.  The initial energy
	of the corresponding electrons in the right lead is far below the Fermi
	surface.  Therefore the creation of holes corresponds to heating the right
	lead, which visible in the
	large positive value of $\dot Q$ at $V_\text{gate}=-5\Gamma$.  For less negative
	$V_\text{gate}$, the energy of the created holes is also less negative and $\dot
	Q$ is diminished.  Once the ground state also lies above the Fermi energy,
	the process comes to rest.  For positive $V_\text{gate}$, the spin-down
	channel is active and adds hot electrons.  While this leads to an opposite
	electric current, it also corresponds to heating.
	
	For zero temperature, the borders between the different regions are
	sharp and both the current and the heat balance between the peaks vanish
	exactly.  For larger temperatures, the curves smear out.  While at
	$V_\text{gate}=0$, the electric current must vanish for symmetry reasons,
	the heat balance assumes negative values (see inset of Fig.~\ref{fig:U0}).
	This means that we find a cooling process in which hot electrons of the right lead
	are replaced by cold ones.
	
	\begin{figure}
		\includegraphics[]{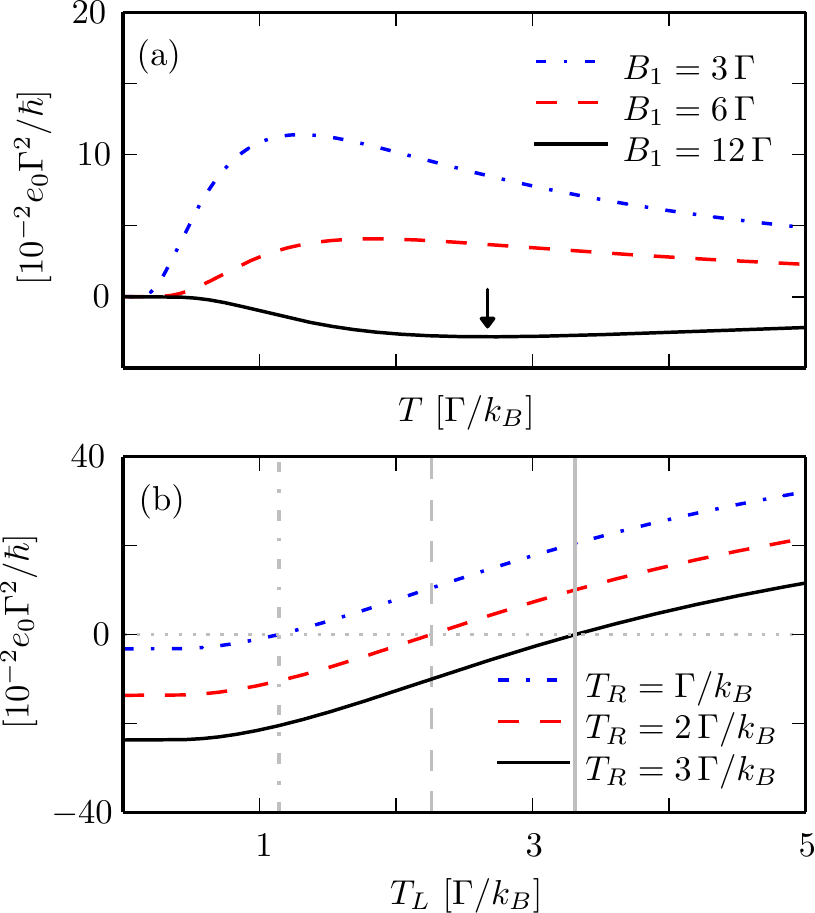}\caption{\label{fig:temperature}
		Heat current as a function of the temperature of the left lead for (a) equal
		temperatures $T=T_L=T_R$ and (b) for pumping against a temperature gradient
		for various $T_R$.
		The Zeeman splitting in dot~2 is $B_2=6\,\Gamma$, while all other parameters
		are as in Fig.~\ref{fig:U0}. The arrow in panel (a) marks
		the minimum of the curve for $B_1=12\,\Gamma$; the vertical lines in panel (b)
		mark the zeros of the heat current and, therewith, the temperature up to which
		cooling can be achieved.
		}
	\end{figure}%
	\begin{figure}
		\includegraphics[]{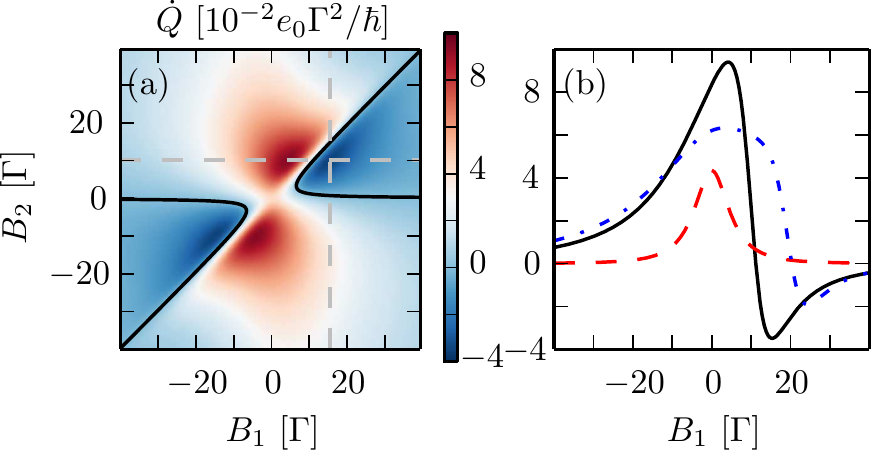}\caption{\label{fig:Bfield}
		(a) Heat production rate on the right lead as a function of the Zeeman splittings $B_1$ and $B_2$ at the
		symmetry point for $T=3\Gamma/k_B$ and other parameters as in Fig.~\ref{fig:U0}. The cross line indicates
		its minimum for positive Zeeman splittings. The solid contour line indicates
		vanishing heat current. (b) Corresponding slices at constant
		$B_2=10\Gamma$ (solid line), $B_2=20\Gamma$ (dashed), and
		$B_2=30\Gamma$ (dashed-dotted).
		}
	\end{figure}%
	
	Figure~\ref{fig:temperature}(a) shows the heat balance as a function of the
	temperature for different Zeeman splittings.  Interestingly, at zero
	temperature, $\dot Q=0$ irrespective of the magnetic field gradient, i.e.,
	not only the cooling but also the heating vanishes as expected.  Moreover, the data
	indicate that cooling is possible only on the side on which the splitting
	is smaller.  In the symmetric situation (dashed curve for $B_1 = B_2$),
	we already observe significant heating in the right lead.  For symmetry
	reasons it equals the heating of the left lead.  Therefore, the coupling to
	the out-of-equilibrium QPC augments the total thermal energy,
	a behavior that agrees with our expectations.
	
	A more systematic investigation of the magnetic field dependence is
	provided in Fig.~\ref{fig:Bfield}.  The overview as a function of $B_1$ and
	$B_2$ indicates that the optimal cooling of the right is found for $B_2$
	clearly smaller than $B_1$, but not too small.  One might have expected
	that $B_1$ should be as large as possible to ensure a strong population of
	the relevant states in the left lead.  However, in such a case the onsite levels are
	strongly detuned so that the effective transition matrix element of the perturbation
	between the DQD eigenstates becomes small; see \ref{sec:analytics}.
	
	A most intriguing question is whether one can reduce the thermal energy of one
	lead even when it is already at a lower temperature than the other lead,
	i.e., whether one can pump heat from cold to hot.  In
	Fig.~\ref{fig:temperature}(b) we show the results for pumping heat against
	a temperature gradient.  Globally, we find that this is possible for
	moderate temperature differences of roughly $10\%$.
	
	\section{Coulomb repulsion}
	
	To make our study applicable to realistic quantum dots, we have to include
	Coulomb repulsion.  For $U=0$, we had chosen our working point at
	$V_\text{gate}=0$ where both channels are symmetric with respect to the
	Fermi energy.  Therefore we start by writing the interaction term in
	the DQD Hamiltonian~\eqref{Hdqd} in a more symmetric form with the help of
	the identity
	\begin{equation}		\label{eq:n-n}
		\begin{split}
			\sum_{\kappa\neq\kappa'} n_\kappa n_{\kappa'}
			= &\sum_{\kappa\neq\kappa'} \Big(n_\kappa-\frac{1}{2}\Big) \Big(n_{\kappa'}-\frac{1}{2}\Big)\\
			&+ (N-1)\sum_{\kappa} n_\kappa -\frac{1}{4}N(N-1) ,
		\end{split}
	\end{equation}
	where $n_\kappa = c^\dagger_\kappa c_\kappa$ is the occupation of the
	single particle level $\kappa = 1,\ldots,N$.  The interpretation of this
	identity is that our interaction Hamiltonian can be expressed by the
	particle-hole symmetric terms $c_\kappa^\dagger c_\kappa-1/2 = -(c_\kappa
	c_\kappa^\dagger-1/2)$ plus an onsite energy.  To be specific, the
	interaction term in Eq.~\eqref{Hdqd} is particle-hole symmetric if a gate
	voltage shifts the $N=4$ levels by $-3U/2$ [recall the prefactor $U/2$ in
	the interaction term of the Hamiltonian (\ref{Hdqd})].
	
	\begin{figure}
		\includegraphics[]{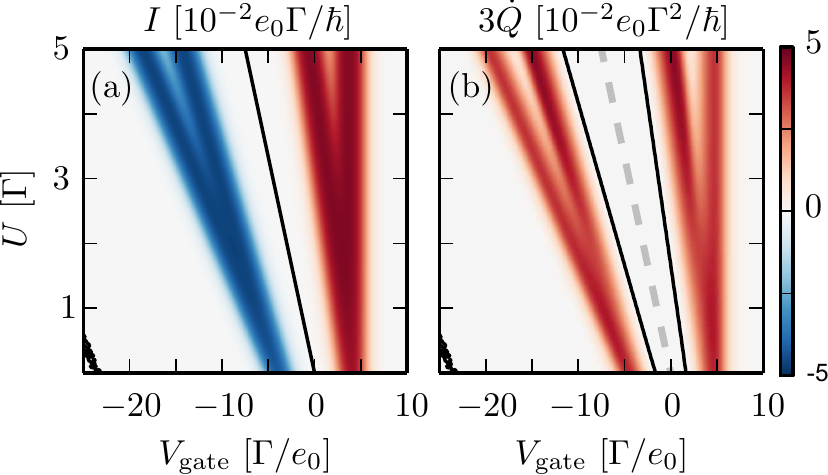}\caption{\label{fig:UV}
		(a) Electric current and (b) heat balance as a function of the gate voltage
		$V_{\textrm{gate}}$ and the Coulomb interaction $U$ for $k_BT=\Gamma/2$.
		Other parameters are as in Fig.~\ref{fig:U0}. Black contour lines
		indicate vanishing current and heat current, respectively. The grey dashed
		line, $V_{\textrm{gate}}=-3U/2$, in panel (b) marks the points for which the electric
		current is zero. That line tracks the particle-hole symmetry point which is displaced by the
		interaction.
		}
	\end{figure}%
	\begin{figure}
		\includegraphics[]{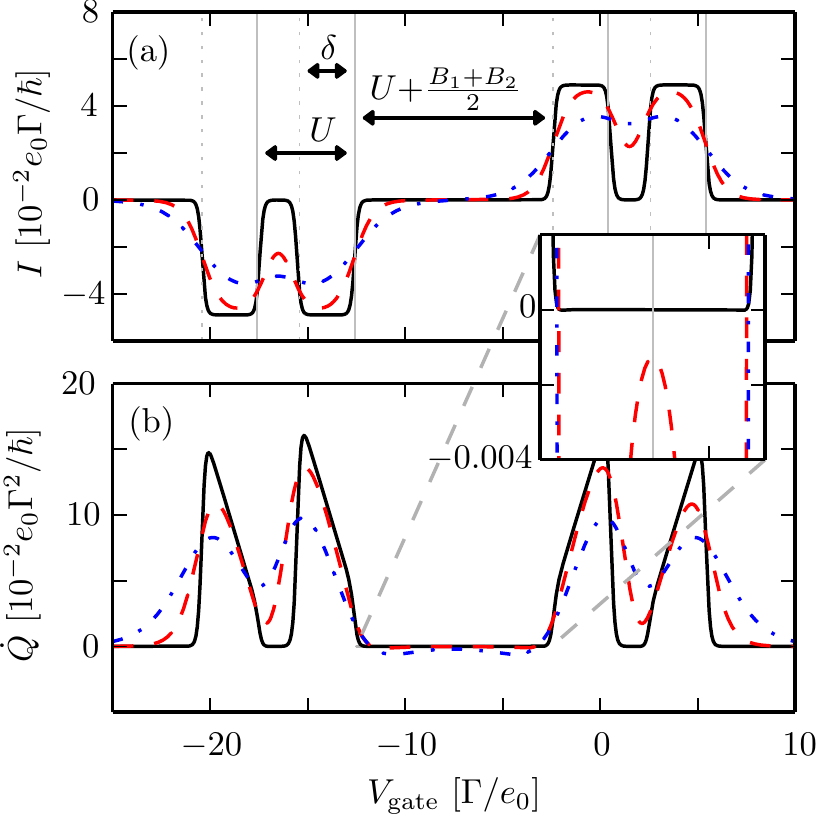}\caption{\label{fig:realistic}
		(a) Current and (b) heat current as function of the gate voltage
		$V_{\textrm{gate}}$ as in Fig.~\ref{fig:U0} but for nonzero
		Coulomb interaction $U=5\Gamma$. Vertical lines indicate the
		transition energies listed in \ref{sec:analytics}. Inset: Heat current in the
		region close to the symmetry point $V_{\textrm{gate}}=-(3/2)U$
		which is indicated by a vertical line.
		}
	\end{figure}%
	
	The predicted shift of the operating point is indeed visible in the current
	and the heat balance as a function of the gate voltage and the interaction
	strength plotted in Fig.~\ref{fig:UV}.  The plot also reveals that with
	increasing interaction, each current peak splits into two peaks with a
	distance $U$.  The values of these shifts can be appreciated in the
	horizontal slices of both panels shown in Fig.~\ref{fig:realistic}.  The
	physical reason for the peak separation is that for each channel, the
	onsite energies are relevant for the empty channel, while for the
	occupation with a further electron, one must overcome the Coulomb
	repulsion.  This effectively augments the excitations energies by $U$.
	
	\begin{figure}
		\includegraphics[]{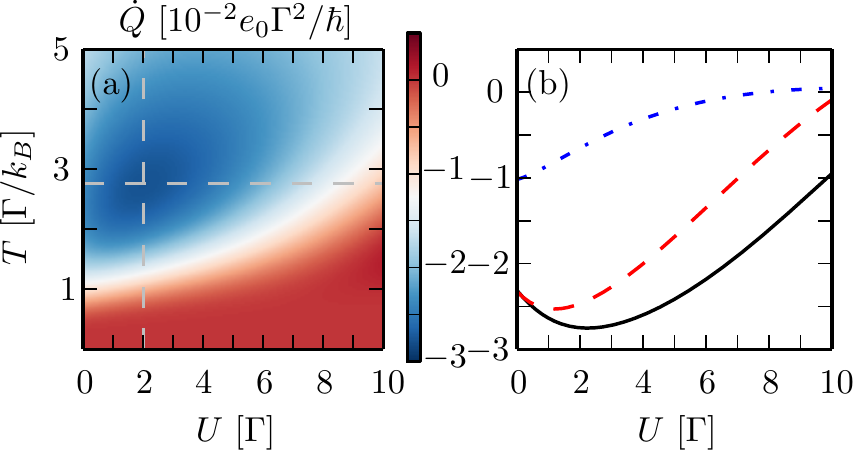}\caption{\label{fig:heatdecay}
		(a) Heat current on the right lead at the symmetry point, as a function of the DQD temperature and
		the Coulomb interaction for parameters as in Fig.~\ref{fig:U0}. The cross line
		indicates its minimum. (b) Corresponding slices at constant
		$T=3\Gamma/k_B$ (solid line), $T=2\Gamma/k_B$ (dashed), and
		$T=\Gamma/k_B$ (dashed-dotted).
		}
	\end{figure}%
	As a drawback, the interaction energy not only shifts the working point,
	but also changes $\dot Q$ quantitatively.  The data in
	Fig.~\ref{fig:heatdecay} shows that the cooling effect has a maximum at
	finite $U$ where the location of the optimum value depends on the temperature.
	Thus, small interactions may be beneficial for cooling, but with increasing
	$U$, the cooling power decays and may even turn into heating.  Thus an
	experimental realization of our scheme should be attempted with quantum
	dots that have a large capacity.
	
	\section{Conclusions}
	We have shown that a double quantum dot subject to an asymmetric Zeeman splitting
	and capacitively coupled to the charge fluctuations of a nearby quantum point contact in the tunnel limit
	can experience rectification. In a suitable region of parameter space, the system can act as a heat
	pump that cools the lead coupled to the dot with the weaker Zeeman splitting, even if that lead
	is already the coldest one. Such a refrigeration scheme can operate with a zero net electric current across
	the double quantum dot system. This is possible thanks to a particle-hole symmetry that can be preserved
	if the gate voltage is suitably adjusted to compensate the effect of Coulomb repulsion.
	For this setup to act as a heat pump, electric current creating shot noise must flow through
	the nearby quantum point contact. However, that electric current does not need to run parallel to the line
	connecting the two heat reservoirs through the double quantum dot system. This feature can become an
	important advantage in potential applications.

	\section*{Acknowledgments}
	
	We would like to thank Rafael S\'anchez for
	valuable comments. This work has been supported by Spain's MINECO under grants no. FIS2013-41716-P and MAT2014-58241-P.
	
	\appendix
	\section{Analytical solution for an individual channel}
	\label{sec:analytics}
	
	Transport through an individual spin channel can be described by a
	Pauli-type rate equation $\dot P = \mathcal{M}P$ for the occupation
	probabilities $P=(P_0,P_g,P_e)^\mathsf{T}$ in the energy basis
	${\ketbra{0}{0}, \ketbra{g}{g}, \ketbra{e}{e}}$. The ground and
	excited states are parametrized \cite{BrandesPR2005a} by
	\begin{equation}
		\ket{g} = -\sin\theta\ket{1} + \cos\theta\ket{2},\quad
		\ket{e} = \cos\theta\ket{1} + \sin\theta\ket{2},	
	\end{equation}
	where $\ket{1}$ and $\ket{2}$ refer to the left and right DQD state,
	respectively, in the local basis. The geometrical factors are determined by
	$\cos(2\theta)=-\epsilon_\sigma/\delta$ and $\sin(2\theta)=\abs{\Omega}/\delta$,
	with level splitting $\delta= \sqrt{\epsilon_\sigma^2+\abs{\Omega}^2}$
	and detuning
	\begin{align}
	\epsilon_\sigma=\frac{B_1-B_2}{2}\sgn(\sigma),
	\end{align}
	where $\sgn(\sigma)$ is positive for $\sigma=\,\uparrow$ and negative, otherwise.
	
	The Liouvillian reads
	\begin{align}
	\mathcal{M} =
			\begin{bmatrix}
		-a -b	& \bar a 			& \bar b\\
		a		& -\bar a -\gamma'	& \gamma'\\
		b		& \gamma'		& -\bar b -\gamma'\\
			\end{bmatrix},
	\end{align}
	where $a = \Gamma -\bar a = \Gamma f(E_g-\mu)$ and $b = \Gamma -\bar b = \Gamma f(E_e-\mu)$
	are finite dot-lead tunneling rates.
	Albeit the leads are in equilibrium, electrons can still be
	excited by the driving of the QPC, which is modeled by
	Fermi's golden rule rate $\gamma' = (s/2)^2\sin^2(2\theta)C(-\delta)$
	of the coupling operator $H_\textrm{QPC}^\textrm{tun}$.
	From the current operator
	\begin{align}
	\mathcal{J} &=
			\begin{bmatrix}
		0						& \bar a \sin^2(\theta)	& \bar b \cos^2(\theta)\\
		-a \sin^2(\theta)		& 0						& 0\\
		-b \cos^2(\theta)		& 0						& 0\\
			\end{bmatrix},
\intertext{and the heat current operator}
	\mathcal{J}_Q &= \frac{\delta}{2}
			\begin{bmatrix}
		0						& \bar a \sin^2(\theta)	& -\bar b \cos^2(\theta)\\
		-a \sin^2(\theta)		& 0						& 0\\
		b \cos^2(\theta)		& 0						& 0\\
			\end{bmatrix},
	\end{align}
	follows directly the current $I=\tr\mathcal{J}P^{\textrm{st}}$ and the heat current
	${\dot Q}=\tr\mathcal{J_Q}P^{\textrm{st}}$, respectively. Hereby, $\tr=(1,1,1)$ denotes the
	trace operator and
	\begin{align}
	P^{\textrm{st}} = \frac{1}{\Gamma^2-ab +(2\Gamma+a+b)\gamma'}
			\begin{bmatrix}
		\bar a \bar b + (\bar a+\bar b)\gamma'\\
		a\bar b +  (a+b)\gamma'\\
		\bar a b + (a+b)\gamma'\\
			\end{bmatrix}
	\end{align}
	the stationary state with $\mathcal{M}P^{\textrm{st}} =0$.
	Finally, we obtain
	\begin{align}
	I 	={}& \gamma'\Gamma\cos(2\theta)\frac{a -b}{
			\Gamma^2 -ab +(2\Gamma +a +b)\gamma'
		} \nonumber\\
		={}& \frac{s^2\Gamma C(-\delta)}{4}
		\frac{b -a}{
			\Gamma^2 -ab +(2\Gamma +a +b)\gamma'
		}\frac{\epsilon_\sigma\abs{\Omega}^2}{\delta^3},
\intertext{and}
	{\dot Q} ={}& -\frac{\delta}{2\cos(2\theta)} I -(\mu -V_{\textrm{gate}})I \nonumber\\
		={}& \frac{s^2\Gamma C(-\delta)}{8}
		\frac{b -a}{
			\Gamma^2 -ab +(2\Gamma +a +b)\gamma'
		}\frac{\abs{\Omega}^2}{\delta}\nonumber\\
		{}&+( V_{\textrm{gate}} -\mu)I.
	\end{align}
	The ground energy $E_g$ and the excited energy $E_e$ occurring in the
	Fermi functions are given by $E_{g/e}=V_{\textrm{gate}}-E_{k,\sigma}^\pm$
	with
	\begin{align}
	E_{k,\sigma}^\pm = - k U \pm \delta/2
			   -\frac{B_1+B_2}{4}\sgn(\sigma).
	\label{eq.:levelEnergies}
	\end{align}
	The latter includes a displacement by a multiple $k$ of the Coulomb interaction.
	To emphasize explicitly the dependence on the spin $\sigma$ and this
	multiple $k$, we write in the following the current of the individual
	channels as $I=I_{k,\sigma}$.
	
	The two-channel case can be approximated by summation of the
	single-channel solutions for spin-up and spin-down. For vanishing
	Coulomb interaction, the two channel current is obtained from
	$I_{0,\downarrow}+I_{0,\uparrow}$, while it is composed of
	$I_{0,\downarrow}+I_{1,\downarrow}+I_{2,\uparrow}+I_{3,\uparrow}$
	for finite $U\gg\delta$. The analytical heat current for the two
	channels case can be analogously defined. The two channels case
	for vanishing Coulomb interaction is in Fig.~\ref{fig:U0} compared
	to the numerical solution of the full master equation.
	Further, figure~\ref{fig:realistic} shows the transition energies for finite
	Coulomb interaction---the vertical lines, from left to
	right, are given by the energies $E_{k,\sigma}^s$ at $(k,\sigma,s)=
	(3,\uparrow,-),(3,\uparrow,+),(2,\uparrow,-),(2,\uparrow,+),
	(1,\downarrow,-),(1,\downarrow,+),(0,\downarrow,-),(0,\downarrow,+)$.

\end{document}